\documentclass[letterpaper,conference]{IEEEtran}

\usepackage[encapsulated]{CJK}
\usepackage{ucs}
\usepackage[utf8x]{inputenc}
\usepackage[cmex10]{amsmath}
\usepackage{amssymb,amscd,bbm,amsthm,mathrsfs,dsfont}
\usepackage{algorithmic,algorithm}
\usepackage{mdwmath}
\usepackage{mdwtab}
\usepackage{bm,upgreek}
\usepackage{cite}
\usepackage{graphicx,psfrag}
\usepackage{array}
\usepackage{booktabs}
\usepackage{indentfirst}
\usepackage{subfigure}
\usepackage{lipsum,fancyhdr,lastpage,refcount}
\usepackage{mathtools}
\usepackage[T1]{fontenc}
\usepackage{url}

\IEEEoverridecommandlockouts

\let\oldnl\nl
\newcommand{\nonl}{\renewcommand{\nl}{\let\nl\oldnl}}

\begin{document}

\title{Performance Optimization in Mobile-Edge Computing via Deep Reinforcement Learning}

\author{\IEEEauthorblockN{Xianfu Chen, Honggang Zhang, Celimuge Wu, Shiwen Mao, Yusheng Ji, and Mehdi Bennis}

\thanks{X. Chen is with the VTT Technical Research Centre of Finland, Finland (e-mail: xianfu.chen@vtt.fi). H. Zhang is with the College of Information Science and Electronic Engineering, Zhejiang University, Hangzhou, China (e-mail: honggangzhang@zju.edu.cn). C. Wu is with the Graduate School of Informatics and Engineering, University of Electro-Communications, Tokyo, Japan (email: clmg@is.uec.ac.jp). S. Mao is with the Department of Electrical and Computer Engineering, Auburn University, Auburn, AL, USA (email: smao@ieee.org). Y. Ji is with the Information Systems Architecture Research Division, National Institute of Informatics, Tokyo, Japan (e-mail: kei@nii.ac.jp). M. Bennis is with the Centre for Wireless Communications, University of Oulu, Finland (email: bennis@ee.oulu.fi).}
}

\maketitle

\begin{abstract}

To improve the quality of computation experience for mobile devices, mobile-edge computing (MEC) is emerging as a promising paradigm by providing computing capabilities within radio access networks in close proximity.
Nevertheless, the design of computation offloading policies for a MEC system remains challenging.
Specifically, whether to execute an arriving computation task at local mobile device or to offload a task for cloud execution should adapt to the environmental dynamics in a smarter manner.
In this paper, we consider MEC for a representative mobile user in an ultra dense network, where one of multiple base stations (BSs) can be selected for computation offloading.
The problem of solving an optimal computation offloading policy is modelled as a Markov decision process, where our objective is to minimize the long-term cost and an offloading decision is made based on the channel qualities between the mobile user and the BSs, the energy queue state as well as the task queue state.
To break the curse of high dimensionality in state space, we propose a deep $Q$-network-based strategic computation offloading algorithm to learn the optimal policy without having a priori knowledge of the dynamic statistics.
Numerical experiments provided in this paper show that our proposed algorithm achieves a significant improvement in average cost compared with baseline policies.

\end{abstract}

\section{Introduction}
\label{intr}

With the proliferation of smart devices, more and more mobile applications, such as location-based virtual/augmented reality and online gaming, are emerging and gaining popularity\cite{Cisc17}.
However, the mobile devices are in general resource-constrained, for example, the battery capacity and the local CPU computation power are limited.
The tension between computation-intensive applications and resource-constrained mobile devices creates a hurdle of having satisfactory Quality-of-Service (QoS) and Quality-of-Experience (QoE), and is hence driving a revolution in terms of computing infrastructure \cite{Saty17}.

Mobile-edge computing (MEC) is envisioned as a promising paradigm to address the hurdle by providing computing capabilities within radio access networks (RANs) in close proximity to mobile users (MUs) \cite{Mach17, Liu17}.
By offloading computation tasks to the resource-rich MEC servers, not only the computation QoS and QoE can be greatly improved, but the capabilities of mobile devices can be augmented for running resource-demanding applications.
Recently, lots of efforts have been centered on the design of computation offloading policy.
In \cite{Wang17}, Wang et al. developed an alternating direction method of multipliers-based algorithm to resolve the issue of revenue maximization by optimizing computation offloading decisions, resource allocation and content caching strategies.
In \cite{Hu18}, Hu et al. proposed a two-phase based method for joint power and time allocation while considering cooperative computation offloading in a wireless power transfer-assisted MEC system.
The policies in these works are primarily based on one-shot optimization and fail to characterize long-term computation offloading performance.

For a MEC system, the computation offloading process requires wireless data transmission, for which the design of computation offloading policies should take into account the existing environmental dynamics, such as the time-varying channel quality and the task arrival and energy status at a mobile device.
In \cite{Liu16}, Liu et al. formulated the problem of delay-optimal computation task offloading under a Markov decision process (MDP) framework and developed an efficient one-dimensional search algorithm to find the optimal solution.
The challenge for the work in \cite{Liu16} lies in the dependence on statistical information of channel quality variations and task arrivals.
In \cite{Mao16}, Mao et al. investigated a dynamic computation offloading policy for a MEC system with wireless energy harvesting-enabled mobile devices using Lyapunov optimization techniques.
However, the Lyapunov optimization can only construct an approximately optimal solution.
Xu et al. developed in \cite{Xu17} a reinforcement learning based algorithm to learn the optimal computation offloading policy, which at the same time does not need a priori knowledge of environmental statistics.

When the MEC meets an ultra dense RAN, a number of base stations (BSs) are available with different data transmission qualities.
In this context, the explosion in state space makes the conventional reinforcement learning algorithms \cite{Xu17, Watk12, Rich98} infeasible.
The focus of this paper is to consider the MEC in an ultra dense system, where the mobile devices are wireless charging enabled.
The problem of designing an optimal computation offloading policy is formulated as a MDP.
We resort to a deep neural network based function approximator to deal with the curse of state space explosion \cite{Mnih15}.
As a major contribution, we propose an online strategic computation offloading policy based on a deep $Q$-network (DQN), with which a typical MU in the ultra dense MEC system is able to realize a significant performance improvement.

The rest of the paper is organized as follows.
In the next section, we describe the system model and the basic assumptions considered in this paper.
In Section \ref{prob}, we formulate the problem of designing an optimal computation offloading policy as a MDP.
We detail the proposed algorithm in Section \ref{solu}.
To validate the proposed study, we provide numerical experiments under various settings in Section \ref{simu}.
Finally, we draw the conclusions in Section \ref{conc}.

\section{System Model and Assumptions}
\label{sysm}

\begin{figure}[t]
  \centering
  \includegraphics[width=18pc]{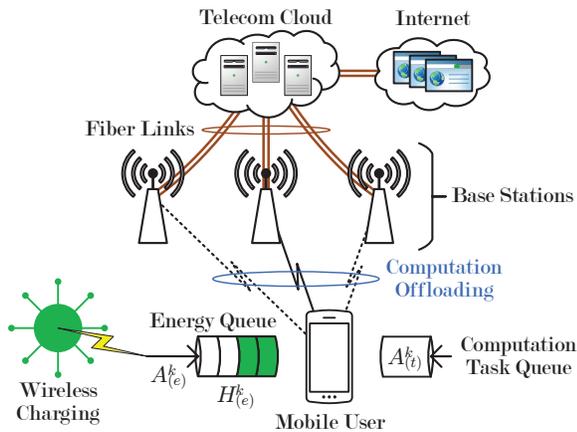}
  \caption{Illustration of a mobile-edge computing system with wireless charging enabled mobile devices.}
  \label{systMode}
\end{figure}

As depicted in Fig. \ref{systMode}, we shall consider in this paper an ultra dense networking environment covered by a set $\mathcal{N} = \{1, \cdots, N\}$ of BSs.
The BSs are connected via the fiber cables to a resource-rich computing infrastructure, namely, the telecom cloud, which is deployed by the network operator.
By strategically offloading the computation tasks for cloud execution, the wireless charging enabled MUs can expect a significantly improved computation experience.
In the analysis that follows, we focus on a representative MU in the dense RAN.
The time horizon is discretized into epochs, each of which is of equal duration $\delta$ (in seconds) and is indexed by an integer $k \in \mathds{N}_+$.
The whole system operates over a common spectrum, and we denote the frequency bandwidth by $W$ (in Hz).

We denote the channel gain state between the MU and a BS $n \in \mathcal{N}$ during each epoch $k$ as $g_n^k$, which independently picks a value from a finite state space $\mathcal{G}_n$.
The channel state transitions across the epochs are modelled as a finite-state discrete-time Markov chain.
We let $\mu$ (in bits) represent the input data size of a computation task.
The computation task arrivals at the MU are assumed to be an independent and identically distributed sequence of Bernoulli random variables with a common parameter $\lambda_{(t)} \in [0, 1]$.
More specifically, we choose $A^k_{(t)} \in \{0, 1\}$ as the task arrival indicator, that is, $A^k_{(t)} = 1$ if a task is generated at the beginning of epoch $k$ and otherwise, $A^k_{(t)} = 0$.
Then, $\textsf{Pr}\{A^k_{(t)} = 1\} = 1 - \textsf{Pr}\{A^k_{(t)} = 0\} = \lambda_{(t)}$, where $\textsf{Pr}\{\cdot\}$ denotes the probability of an event.
In our considered MEC system, the computation task can be either executed locally at the mobile device of the MU or offloaded to and executed at the telecom cloud.
At the beginning of each epoch $k$, the MU makes a joint decision regarding computation offloading $T^k \in \{-1\} \cup \{0\} \cup \mathcal{N}$ and energy allocation $E^k$ (in energy units).
Note that $T^k = -1$ is the case the MU decides not to execute the computation task, and there will be no computation task execution delay and $E^k = 0$, hence leading to the dropping of an arrived task.

We have $T^k = 0$ if the computation task is executed locally at the mobile device of the MU during an epoch $k$.
Let $\nu$ be the number of CPU cycles required to process one input bit of the computation task.
Then the allocated CPU-cycle frequency at the MU can be calculated as
\begin{align}\label{cyclFreq}
  f^k_{(l)} = \sqrt{\frac{E^k}{\tau \mu \nu}},
\end{align}
with the given energy $E^k$, where $\tau$ is the effective switched capacitance that depends on chip architecture of the mobile device \cite{Burd96}.
Moreover, the CPU-cycle frequency is constrained by $f_{(l)}^k \leq \overline{f}$.
The incurred delay for local computation execution at epoch $k$ can be hence expressed by
\begin{align}\label{locaDela}
  b^k_{(l)} = \frac{\mu \nu}{f^k_{(l)}}.
\end{align}

At the beginning of epoch $k$, if the MU decides to offload the computation task to the telecom cloud for execution via a BS $n \in \mathcal{N}$, namely, $T^k = n$, the input data should be first transmitted to the cloud.
In a dense networking scenario, the achievable data rate can be written as
\begin{align}\label{dataRate}
  R^k = W \log_2\!\!\left(1 + I^{-1} g^k_n \frac{E^k}{b^k_{(c), (tr)}}\right),
\end{align}
where $I$ is the received average power of interference and additive background noise, while $b^k_{(c),(tr)}$ is the transmission time and is the solution of
\begin{align}\label{tranTime}
  R^k b^k_{(c),(tr)} = \mu.
\end{align}
In (\ref{tranTime}) above, we suppose that the energy $E^k$ is evenly assigned to the input bits of the computation task \cite{Yang12}.
After receiving the input bits of the offloaded computation task, the telecom cloud proceeds to execute it.
We let $f_{(c)}$ be the constant CPU-cycle frequency assigned to the MU, which is based on the subscribed cloud-computing contract between the MU and the network operator.
The execution time of the computation task at the cloud takes up to
\begin{align}\label{execTime}
  b_{(c), (ex)} = \frac{\mu \nu}{f_{(c)}}.
\end{align}
Therefore, the overall delay resulted from offloading computation task for cloud execution is
\begin{align}\label{clouDela}
  b^k_{(c)} = b^k_{(c),(tr)} + b_{(c), (ex)},
\end{align}
where as in the existing works \cite{Mao16, Chen16}, we neglect the time consumed for sending the computation outcomes from the telecom cloud back to the MU.

Let $H^k$ be the energy queue length of the MU at the beginning of epoch $k$, which evolves according to
%
%Following the above discussions, the task queue evolves according to
%
%\begin{align}\label{taskQueu}
%   H^{k + 1}_{(t)} = \min\!\left\{H^k_{(t)} - \mathds{1}_{\left\{T^k \neq -1\right\}} + A^k_{(t)}, \overline{H}_{(t)}\right\},
%\end{align}
%
%where $\overline{H}_{(t)}$ is the maximum number of computation tasks that can be queued at the MU and $\mathds{1}_{\{\Omega\}}$ denotes the indicator function that equals 1 if condition $\Omega$ is satisfied and otherwise, 0.
%
%The energy queue evolves according to
%
\begin{align}\label{enerQueu}
  H^{k + 1} = \min\!\left\{H^k - E^k + A^k_{(e)}, \overline{H}\right\},
\end{align}
where $\overline{H}$ limits the maximum number of energy units that can be stored and $A^k_{(e)}$ is the number of energy units acquired from the wireless environment by the end of epoch $k$.

\section{Problem Formulation}
\label{prob}

The computation task arrivals from the MU can be offloaded to the telecom cloud depending on the channel qualities, the energy queue state and the computation task queue state.
We denote $\mathbf{x}^k = (A^k_{(t)}, H^k, \mathbf{g}^k) \in \mathcal{X} = \{0, 1\} \times \{0, 1, \cdots, \overline{H}\} \times \{\times_{n \in \mathcal{N}} \mathcal{G}_n\}$ as the network state of the MU at each epoch $k$, where $\mathbf{g}^k = (g^k_n, n \in \mathcal{N})$.
With observation $\mathbf{x}^k$ at the beginning of epoch $k$, the MU strategically decides an action $\mathbf{y}^k = (T^k, E^k) \in \mathcal{Y} = \{\{-1\} \cup \{0\} \cup \mathcal{N}\} \times \{0, 1, \cdots, \overline{H}\}$ following a stationary control policy $\bm\Phi = (\Phi_{(t)}, \Phi_{(e)})$, where $\Phi_{(t)}$ and $\Phi_{(e)}$ are, respectively, the computation offloading and the energy allocation policies.
That is, $\bm\Phi(\mathbf{x}^k) = (\Phi_{(t)}(\mathbf{x}^k), \Phi_{(e)}(\mathbf{x}^k)) = (T^k, E^k)$.
Given $\bm\Phi$, the $\{\mathbf{x}^k: k \in \mathds{N}_+\}$ is a controlled Markov chain with the state transition probability as below,
\begin{align}\label{statTran}
     \textsf{Pr}\!\left\{\mathbf{x}^{k + 1} | \mathbf{x}^k, \bm\Phi\!\left(\mathbf{x}^k\right)\right\}
 & = \left(\prod\limits_{n \in \mathcal{N}} \textsf{Pr}\!\left\{g^{k + 1}_n | g^k_n\right\}\right) \textsf{Pr}\!\left\{A^{k + 1}_{(t)}\right\}  \nonumber\\
 & \times \textsf{Pr}\!\left\{H^{k + 1} | H^k, \bm\Phi\!\left(\mathbf{x}^k\right)\right\}.
\end{align}

When the MU is associated with a BS at epoch $k$, which is different from the previous one, additional handover delay is incurred.
We assume that the energy consumption during the handover procedure is negligible for the MU and the delay during the occurrence of one handover is $\zeta$ (in seconds).
Then the handover delay $b^k_{(h)} = b_{(h)}(\mathbf{x}^k, \mathbf{y}^k)$ at epoch $k$ is
\begin{align}\label{handDela}
  b_{(h)}\!\left(\mathbf{x}^k, \mathbf{y}^k\right) = \zeta
  \mathds{1}_{\left\{\left\langle T^k \in \mathcal{N}\right\rangle \wedge \left\langle T^j \in \mathcal{N}\right\rangle \wedge \left\langle T^k \neq T^j\right\rangle\right\}},
\end{align}
where $\mathds{1}_{\{\Omega\}}$ is an indicator function that equals 1 if condition $\Omega$ is met and otherwise, 0, and $j = \max\{\ell: T^{\ell} \in \mathcal{N}, \ell \in \mathds{N}_+, \ell < k\}$.
The experienced delay is the key performance indicator for evaluating the quality of a task computing experience.
In addition, due to the sporadic nature of energy units that can be received across the epochs, the newly arriving computation tasks at an epoch may have to be dropped, the cost $b_{(d)}^k = b_{(d)}(\mathbf{x}^k, \mathbf{y}^k)$ of which is defined to be
\begin{align}\label{costDrop}
  b_{(d)}\!\left(\mathbf{x}^k, \mathbf{y}^k\right) = \textsf{Pr}\!\left\{A^k_{(t)} = 1\right\} \mathds{1}_{\{T^k = -1\}}.
\end{align}
In line with the discussions in previous sections, we define the task execution cost $p^k = p(\mathbf{x}^k, \mathbf{y}^k)$ at each epoch $k$ as the weighted sum of the execution delay, the handover delay and the computation task dropping cost, namely,
\begin{align}\label{costFunc}
   p\!\left(\mathbf{x}^k, \mathbf{y}^k\right) = b\!\left(\mathbf{x}^k, \mathbf{y}^k\right) + \rho b^k_{(h)} + \varphi b^k_{(d)},
\end{align}
where $\rho$, $\varphi \in \mathds{R}_+$ are the weights of the handover delay and the computation task dropping cost, respectively, and
\begin{align}
  b\!\left(\mathbf{x}^k, \mathbf{y}^k\right) =
  \left\{\!\!
  \begin{array}{ll}
     0,          & \mbox{if } T^k = -1; \\
     b^k_{(l)},  & \mbox{if } T^k = 0;  \\
     b^k_{(c)},  & \mbox{otherwise}.
  \end{array}
  \right.
\end{align}

Taking expectation with respect to the per-epoch task execution costs over the randomized network states $\mathbf{x}^k$ and the actions $\mathbf{y}^k$ induced by a given control policy $\bm\Phi$, the expected long-term cost of the MU conditioned on an initial network state $\mathbf{x}^1$ can be expressed as
\begin{align}\label{expeCost}
   V(\mathbf{x}, \bm\Phi) =
   \textsf{E}_{\bm\Phi}\!\!\left[(1 - \gamma) \sum_{k = 1}^\infty (\gamma)^{k - 1} p^k | \mathbf{x}^1 = \mathbf{x}\right],
\end{align}
where $\mathbf{x} = (A_{(t)}, H, \mathbf{g})$ with $\mathbf{g} = (g_n: n \in \mathcal{N})$, $\gamma \in [0, 1)$ is the discount factor, and $(\gamma)^{k-1}$ denotes the discount factor to the $(k-1)$-th power.
The objective of the MU is to design an optimal control policy $\bm\Phi^* = (\Phi_{(t)}^*, \Phi_{(e)}^*)$ that minimizes $V(\mathbf{x}, \bm\Phi)$, for any given initial network state $\mathbf{x}$, which can be formally formulated as
\begin{align}\label{optiPoli}
  \bm\Phi^* = \underset{\bm\Phi}{\arg\min}~V(\mathbf{x}, \bm\Phi), \forall \mathbf{x} \in \mathcal{X}.
\end{align}
We denote $V(\mathbf{x}) = V(\mathbf{x}, \bm\Phi^*)$ as the optimal state-value function, $\forall \mathbf{x} \in \mathcal{X}$.

\section{Solving the Optimal Control Policy}
\label{solu}

The formulated computation offloading optimization in (\ref{optiPoli}) is in essential a single-agent infinite-horizon MDP with the discounted cost criterion.
In this section, we shall first investigate the optimal solution within the conventional MDP framework and then proceed to propose a deep reinforcement learning based scheme with limited network statistics information.

\subsection{Optimal MDP Solution}

The optimal state-value function, namely, $V(\mathbf{x})$, $\forall \mathbf{x} \in \mathcal{X}$, can be achieved by solving the Bellman's optimality equation as in the following lemma \cite{Rich98}.

\noindent\textbf{Lemma 1.}
The optimal state-value function $\{V(\mathbf{x}), \forall \mathbf{x} \in \mathcal{X}\}$ satisfies the Bellman's optimality equation, that is, $\forall \mathbf{x}$,
\begin{align}\label{BellEqua}
  & V(\mathbf{x}) =                                                                                                         \nonumber\\
  & \min_{\mathbf{y} \in \mathcal{Y}}\!\left\{(1 - \gamma) p(\mathbf{x}, \mathbf{y}) +
    \gamma \sum_{\mathbf{x}' \in \mathcal{X}} \textsf{Pr}\{\mathbf{x}' | \mathbf{x}, \mathbf{y}\} V(\mathbf{x}')\right\},
\end{align}
where $p(\mathbf{x}, \mathbf{y})$ is the task execution cost when action $\mathbf{y}$ is performed under network state $\mathbf{x}$ and $\mathbf{x}' = (A_{(t)}', H', \mathbf{g}')$ is the subsequent network state with $\mathbf{g}' = (g_n': n \in \mathcal{N})$.

\textit{Remark 1:}
The size $X$ of the network state space $\mathcal{X}$ can be calculated as $X = 2 \times (1 + \overline{H}) \times \prod_{n \in \mathcal{N}} |\mathcal{G}_n|$, where $|\mathcal{G}|$ means the cardinality of the set $\mathcal{G}$.
It can be observed that $X$ grows exponentially as the number $N$ of BSs increases.

\textit{Remark 2:}
The traditional solutions to (\ref{BellEqua}) are based on the value or the policy iteration \cite{Rich98}, which not only need \emph{complete knowledge} of the channel state transition probabilities, the computation task arrival and the received energy unit statistics but suffer from \emph{exponential computation complexity} due to the extremely huge network state space even with a reasonable number of BSs.
Suppose there is a MEC system with $6$ BSs and for each BS, the channel gain is quantized into $8$ states.
If we set $\overline{H} = 4$ as in the numerical experiments, there are astonishing $2621440$ local network states in total for the MU.

The next subsection thereby focuses on developing a practically efficient scheme to approach the optimal policy.

\subsection{Deep Reinforcement Learning}

\begin{figure}[t]
  \centering
  \includegraphics[width=19pc]{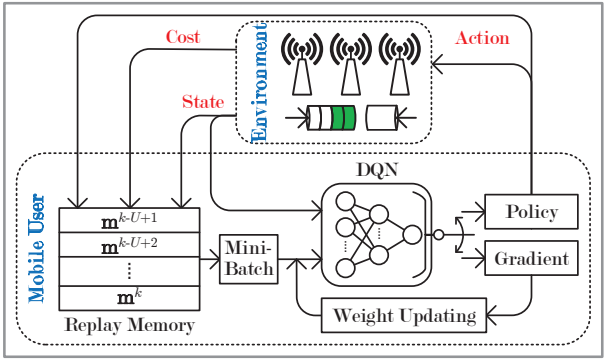}
  \caption{Deep $Q$-network (DQN) based mobile-edge computing system.}
  \label{deepLear}
\end{figure}

Define the right-hand side of (\ref{BellEqua}) as the optimal action-value function $Q: \mathcal{X} \times \mathcal{Y} \rightarrow \mathds{R}$, which is
\begin{align}\label{QFunc}
  Q(\mathbf{x}, \mathbf{y}) = (1 - \gamma) p(\mathbf{x}, \mathbf{y}) + \gamma \sum_{\mathbf{x}' \in \mathcal{X}} \textsf{Pr}\{\mathbf{x}' | \mathbf{x}, \mathbf{y}\} V(\mathbf{x}'),
\end{align}
$\forall (\mathbf{x}, \mathbf{y}) \in \mathcal{X} \times \mathcal{Y}$, we have $V(\mathbf{x}) = \min_{\mathbf{y} \in \mathcal{Y}} Q(\mathbf{x}, \mathbf{y})$, $\forall \mathbf{x} \in \mathcal{X}$.
To address the first technical challenge in Remark 2, we adopt a model-free reinforcement learning scheme called $Q$-learning \cite{Watk12}, which allows us to learn the optimal control policy without any information of dynamic network statistics.

The $Q$-learning scheme is a simple $Q$-function update step, which is performed at the beginning of an epoch.
Based on the observations of the network state $\mathbf{x}^k$, the action $\mathbf{y}^k$, the received task execution cost $p(\mathbf{x}^k, \mathbf{y}^k)$, the computation task arrival $A^{k + 1}_{(t)}$, the number of received energy units $A^k_{(e)}$ at each epoch $k$, and the resulting network state $\mathbf{x}^{k + 1}$ at the next epoch $k + 1$, the MU updates $Q$-function on-the-fly,
\begin{align}\label{QLear}
 & Q\!\left(\mathbf{x}^k, \mathbf{y}^k\right) \leftarrow Q\!\left(\mathbf{x}^k, \mathbf{y}^k\right) + \\
 & \alpha^k \left((1 - \gamma) p\!\left(\mathbf{x}^k, \mathbf{y}^k\right) + \gamma \min_{\mathbf{y} \in \mathcal{Y}}Q\!\left(\mathbf{x}^{k + 1}, \mathbf{y}\right) - Q\!\left(\mathbf{x}^k, \mathbf{y}^k\right)\right), \nonumber
\end{align}
where $\alpha^k \in [0, 1)$ is a time-varying learning rate.
It has been proven that if 1) the network state transition probability under the optimal stationary control policy is stationary, 2) $\sum_{k = 1}^\infty \alpha^k$ is infinite and $\sum_{k = 1}^\infty (\alpha^k)^2$ is finite, and 3) all state-action pairs are visited infinitely often, the convergence of the $Q$-learning process is ensured \cite{Watk12}.
The last condition can be satisfied if the probability of choosing any action in any network state is non-zero (i.e., \emph{exploration}).
Meanwhile, the MU has to exploit the current knowledge in order to perform well (i.e., \emph{exploitation}).
A classical way to balance the trade-off between \emph{exploration} and \emph{exploitation} is the $\epsilon$-greedy strategy \cite{Rich98}.

\textit{Remark 3:}
The $Q$-learning rule, which is formulated in (\ref{QLear}), relieves the dependence on full network statistics information, but still has to face the curse of a huge network state space, as pointed out in Remark 2.

Hereinafter, we adopt a DQN to online estimate the $Q$-function \cite{Mnih15}.
That is, $Q(\mathbf{x}, \mathbf{y}) \approx Q(\mathbf{x}, \mathbf{y}; \bm\theta)$, where $(\mathbf{x}, \mathbf{y}) \in \mathcal{X} \times \mathcal{Y}$ and the set of weights is denoted by $\bm\theta$.
The proposed DQN-based strategic computation offloading for our considered MEC system is illustrated in Fig. \ref{deepLear}.
The MU utilizes a replay memory of a finite size $U$ to store the transition $\mathbf{m}^k = (\mathbf{x}^k, \mathbf{y}^k, p(\mathbf{x}^k, \mathbf{y}^k), \mathbf{x}^{k + 1})$ that is happened at the end of each epoch $k$.
The memory pool is characterized by $\mathcal{O}^k = \{\mathbf{m}^{k - U + 1}, \cdots, \mathbf{m}^k\}$.
According to the experience replay technique, the MU randomly samples an experience at each epoch $k$, i.e., a mini-batch $\tilde{\mathcal{O}}^k \subseteq \mathcal{O}^k$ of $S$ transitions, from $\mathcal{O}^k$ to train the DQN in the direction of minimizing the loss function in (\ref{lossFunc}),
\begin{figure*}
\begin{align}\label{lossFunc}
  L\!\left(\bm\theta^{k + 1}\right) =
  \textsf{E}_{(\mathbf{x}, \mathbf{y}, p(\mathbf{x}, \mathbf{y}), \mathbf{x}')\in \tilde{\mathcal{O}}^k}\!\!\left[\left((1 - \gamma) p(\mathbf{x}, \mathbf{y}) + \gamma Q\!\left(\mathbf{x}', \underset{\mathbf{y}' \in \mathcal{Y}}{\arg\min} Q\!\left(\mathbf{x}', \mathbf{y}'; \hat{\bm\theta}^k\right); \bm\theta^k\right) - Q\!\left(\mathbf{x}, \mathbf{y}; \bm\theta^{k + 1}\right)\right)^2\right]
\end{align}
\hrule
\end{figure*}
where the set of weights of the DQN at an epoch $k$ is denoted as $\bm\theta^k$, and $\hat{\bm\theta}^k$ is a second set of weights for evaluating the action values and is updated to be $\bm\theta^k$ in the next epoch.
The gradient of (\ref{lossFunc}) is given by (\ref{grad}).
\begin{figure*}
\begin{align}\label{grad}
  & \nabla_{\bm\theta^{k + 1}} L\!\left(\bm\theta^{k + 1}\right) = \nonumber\\
  & \textsf{E}_{(\mathbf{x}, \mathbf{y}, p(\mathbf{x}, \mathbf{y}), \mathbf{x}')\in \tilde{\mathcal{O}}^k}\!\!
    \left[\left((1 - \gamma) p(\mathbf{x}, \mathbf{y}) + \gamma Q\!\left(\mathbf{x}', \underset{\mathbf{y}' \in \mathcal{Y}}{\arg\min} Q\!\left(\mathbf{x}', \mathbf{y}'; \hat{\bm\theta}^k\right); \bm\theta^k\right) - Q\!\left(\mathbf{x}, \mathbf{y}; \bm\theta^{k + 1}\right)\right) \nabla_{\bm\theta^{k + 1}} Q\!\left(\mathbf{x}, \mathbf{y}; \bm\theta^{k + 1}\right)\right]
\end{align}
\hrule
\end{figure*}
Algorithm \ref{algo} summarizes the DQN-based online strategic computation offloading for the MU in a MEC system.
\begin{algorithm}[t]
    \caption{DQN-based Online Strategic Computation Task Offloading}
    \label{algo}
    \begin{algorithmic}[1]
        \STATE \textbf{initialize} the replay memory $\mathcal{O}^k$ with a size of $U$, the mini-batch $\tilde{\mathcal{O}}^k$ with a size of $S$, and the $Q$-function with two sets $\bm\theta^k$ and $\hat{\bm\theta}^k$ of random weights, for $k = 1$.

        \REPEAT
            \STATE At the beginning of epoch $k$, observe the network state $\mathbf{x}^k \in \mathcal{X}$ and select an action $\mathbf{y}^k \in \mathcal{Y}$ randomly with probability $\epsilon$ or $\mathbf{y}^k = \arg\min_{\mathbf{y} \in \mathcal{Y}} Q(\mathbf{x}^k, \mathbf{y}; \bm\theta^k)$ with probability $1 - \epsilon$.

            \STATE After deploying $\mathbf{y}^k$, observe the cost $p(\mathbf{x}^k, \mathbf{y}^k)$ and the new network state $\mathbf{x}^{k + 1} \in \mathcal{X}$.

            \STATE Store $\mathbf{m}^k = (\mathbf{x}^k, \mathbf{y}^k, p(\mathbf{x}^k, \mathbf{y}^k), \mathbf{x}^{k + 1})$ in $\mathcal{O}^k$.

            \STATE Sample a random mini-batch of transitions $\tilde{\mathcal{O}}^k \subseteq \mathcal{O}^k$.

            \STATE Update $\bm\theta^{k + 1}$ with the gradient given by (\ref{grad}).

            \STATE Regularly perform $\hat{\bm\theta}^{k + 1} = \bm\theta^k$.

            \STATE Update epoch index by $k \leftarrow k+1$.
        \UNTIL{A predefined stopping condition is satisfied.}
    \end{algorithmic}
\end{algorithm}

\section{Numerical Experiments}
\label{simu}

In this section, we proceed to quantify the performance from our proposed DQN-based online strategic computation offloading.

\subsection{General Setup}

For the DQN, the replay memory is assumed to have a capacity of $U = 5000$ and we select the size of the mini-batch as $S = 100$.
Throughout the numerical experiments, we suppose there are $N = 6$ BSs in the MEC system connecting the MU with the telecom cloud.
The channel gain states between the MU and the BSs are from a common finite set $\{-18, -16, -14, -12,$ $-10, -8, -6, -4\}$ (dB), the transitions of which happen across the epochs following respective randomly generated matrices.
Each energy unit corresponds to $5\times 10^{-5}$ J, and the energy units harvested from the wireless environment follow a Poisson arrival process with average arrival rate $\lambda_{(e)}$.
We set $\rho \zeta = 0.9 \delta$ and $\beta = 9 \delta$.
In addition, $\gamma = 0.9$, $W = 10^6$ Hz, $I = 10^{-3}$ W, $\delta = 10^{-3}$ second, $\overline{H} = 4$, $\mu = 10^3$ bits, $\tau = 10^{-28}$, $\nu = 600$ cycles per bit, $\overline{f} = 1.9$ GHz, and $f_{(c)} = 3.9$ GHz.
For comparisons, we simulate three baselines as well, namely,
\begin{enumerate}
  \item \emph{Local} -- Whenever a computation task arrives, the MU executes it at the local mobile device using the queued energy units.
  \item \emph{Cloud} -- All arriving computation tasks are offloaded to the telecom cloud for computing via the BSs with the best channel qualities.
  \item \emph{Greedy} -- When the computation task queue as well as the energy queue are not empty at an epoch $k$, the MU decides to execute the task locally or at the cloud to achieve the minimum current delay, i.e., $\min\{b_{(l)}^k, b_{(c)}^k\}$.
\end{enumerate}

\subsection{Experimental Results}

We carry out numerical experiments under various settings to validate the proposed work.

\subsubsection{Experiment 1 -- Convergence performance}

\begin{figure}[t]
  \centering
  \includegraphics[width=18pc]{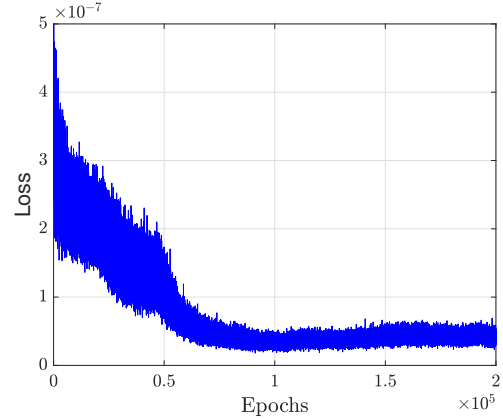}
  \caption{Illustration of convergence property of our proposed algorithm.}
  \label{conver}
\end{figure}

In this experiment, the goal is to validate the convergence property of our proposed DQN-based online computation task offloading algorithm.
We set $\lambda_{(t)} = 0.6$ and $\lambda_{(e)} = 0.5$.
The DQN consists of one hidden layer of 128 neurons.
In Fig. \ref{conver}, we plot the simulated variations in the loss function defined as in (\ref{lossFunc}), which reveals that the convergence of our proposed algorithm can be ensured.
Based on the convergence of loss function, each result in the following experiments is obtained from one system configuration running for $9 \times 10^5$ epochs.

\subsubsection{Experiment 2 -- Performance under different DQN structures}

\begin{figure}[t]
  \centering
  \includegraphics[width=18pc]{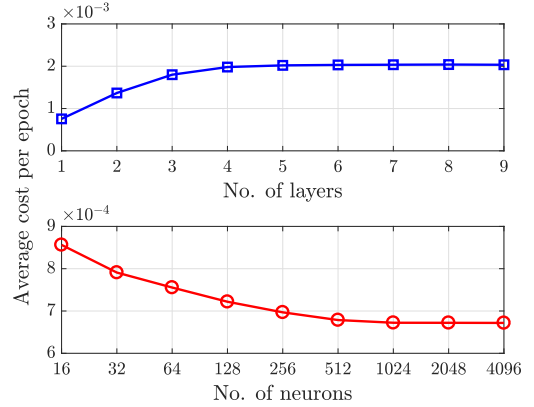}
  \caption{Average cost per epoch versus numbers of layers and neurons.}
  \label{ddqn}
\end{figure}

This experiment tries to demonstrate the MEC performance for the MU in terms of the average cost per epoch using a DQN with different numbers of layers and neurons structures.
We choose $\lambda_{(t)} = 0.4$ and $\lambda_{(e)} = 0.8$ in simulations for the MU.
The results are exhibited in Fig. \ref{ddqn}.
In the upper plot, the number of neurons per hidden layer is fixed to be 64.
It can be observed that a deeper DQN leads to worse average cost performance.
The reason is that over a limited time horizon, adding more hidden layers to the DQN leads to higher training errors \cite{He16}.
In the lower plot, only one hidden layer is implemented in the DQN.
From the curve, better performance is achieved with a bigger number of neurons.
In our considered MEC scenario, a wider (NOT deeper) DQN can better approximate the $Q$-function.

\subsubsection{Experiment 3 -- Performance with changing $\lambda_{(t)}$ and $\lambda_{(e)}$}

\begin{figure}[t]
  \centering
  \includegraphics[width=18pc]{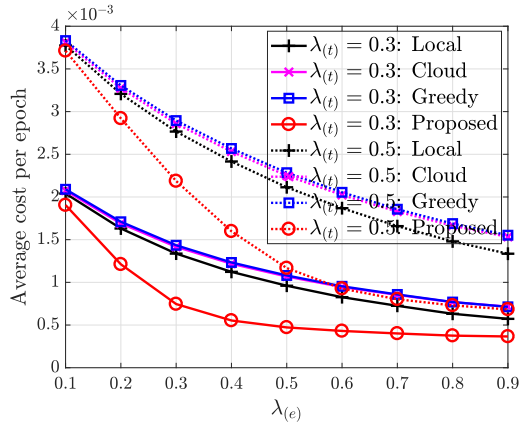}
  \caption{Average cost per epoch versus average energy unit arrival rate.}
  \label{simu01}
\end{figure}
\begin{figure}[t]
  \centering
  \includegraphics[width=18pc]{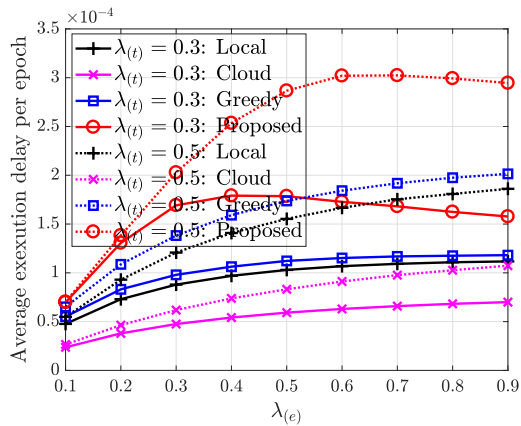}
  \caption{Average execution delay per epoch versus average energy unit arrival rate.}
  \label{simu02}
\end{figure}
\begin{figure}[t]
  \centering
  \includegraphics[width=18pc]{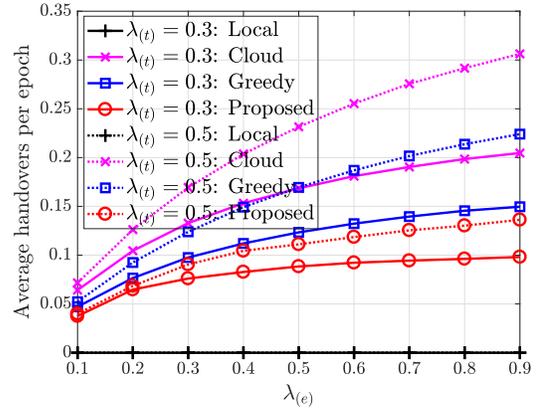}
  \caption{Average handovers per epoch versus average energy unit arrival rate.}
  \label{simu03}
\end{figure}
\begin{figure}[t]
  \centering
  \includegraphics[width=18pc]{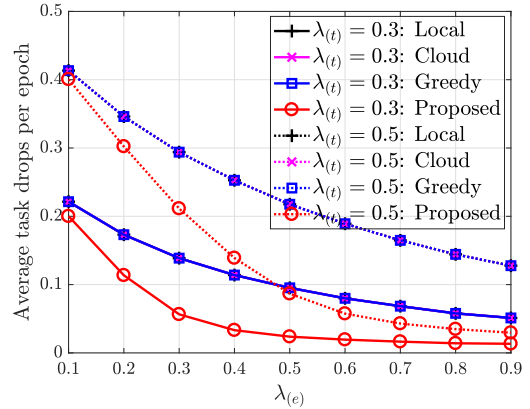}
  \caption{Average task drops per epoch versus average energy unit arrival rate.}
  \label{simu04}
\end{figure}

We do this experiment to simulate the average performance achieved from the proposed DQN-based algorithm and other three baselines versus the average energy arrival rates.
With the findings from Experiment 2, we configure a DQN of one hidden layer with 512 neurons.
The per epoch averages of cost, execution delay, handovers and task drops under $\lambda_{(t)} = 0.3$ and $\lambda_{(t)} = 0.5$ across the entire learning period are depicted in Figs. \ref{simu01}, \ref{simu02}, \ref{simu03} and \ref{simu04}.
From Fig. \ref{simu01}, we can clearly see that compared to the baselines, our proposed algorithm achieves a significant performance improvement in average cost, up to 56\%.
A higher task arriving probability indicates a longer average delay for executing more computation tasks, more handovers between BSs and more task drops, hence a higher average cost.
As the number of energy arrivals increases, the average cost decreases.
This is a result of less computation task drops, which dominate the cost function for the weight choices.
Interestingly, the increase in energy arrivals does not necessarily reduce the task execution delay and the handovers.
This can be explained by the fact that more energy arrivals provide more opportunities for the MU to select a BS with better channel gain to execute a computation task, rather than simply drop it.

\section{Conclusions}
\label{conc}

In this paper, we put our emphasis on investigating the design of a smart computation offloading policy for a MU in an ultra dense network by taking into account the dynamics generated from time-varying channel qualities between the MU and the BSs, harvested energy units and task arrivals.
To solve the formulated MDP, we propose a DQN-based online strategic computation offloading algorithm that survives the curse of high dimensionality in state space and needs no a priori information of dynamics statistics.
We find from numerical experiments that compared to three baselines, our proposed algorithm can achieve minimum long-term cost, up to 56\% in performance improvement, which indicates an optimal tradeoff among the computation task execution delay, the handover delay and the task dropping cost.

\end{document}